\documentclass[11pt]{article}
\usepackage{amsmath,epsf,amssymb,latexsym}

\renewcommand{\theequation}{\thesection.\arabic{equation}}
\newcounter{saveeqn}
\newcommand{\add}{\addtocounter{equation}{1}}
\newcommand{\alpheqn}{\setcounter{saveeqn}{\value{equation}}%
\setcounter{equation}{0}%
\renewcommand{\theequation}{\mbox{\thesection.\arabic{saveeqn}{\alph{equation}}}}}
\newcommand{\reseteqn}{\setcounter{equation}{\value{saveeqn}}%
\renewcommand{\theequation}{\thesection.\arabic{equation}}}

\def\N{$\cal N$}
\def\E {$E_{7(7)}$}
\newif\iffigs\figstrue

\DeclareFontFamily{U}{rsf}{}
\DeclareFontShape{U}{rsf}{m}{n}{
  <5> <6> rsfs5 <7> <8> <9> rsfs7 <10-> rsfs10}{}
\DeclareMathAlphabet\Scr{U}{rsf}{m}{n}

\def\pplogo{\vbox{\kern-\headheight\kern -29pt
\halign{##&##\hfil\cr&{
\ppnumber}\cr\rule{0pt}{2.5ex}&\ppdate\cr}
}}
\makeatletter
\def\ps@firstpage{\ps@empty \def\@oddhead{\hss\pplogo}%
  \let\@evenhead\@oddhead 
}
\def\maketitle{\par
 \begingroup
 \def\thefootnote{\fnsymbol{footnote}}
 \def\@makefnmark{\hbox{$^{\@thefnmark}$\hss}}
 \if@twocolumn
 \twocolumn[\@maketitle]
 \else \newpage
 \global\@topnum\z@ \@maketitle \fi\thispagestyle{firstpage}\@thanks
 \endgroup
 \setcounter{footnote}{0}
 \let\maketitle\relax
 \let\@maketitle\relax
 \gdef\@thanks{}\gdef\@author{}\gdef\@title{}\let\thanks\relax}
\makeatother

\def\dim{{\rm dim}}
\def\iin{{\rm in}}
\def\out{{\rm out}}

\def\slash#1{{\ooalign{\hfil/\hfil\crcr$#1$}}}

\newcommand{\bea}{\begin{eqnarray}}
\newcommand{\eea}{\end{eqnarray}}
\newcommand{\be}{\begin{equation}}
\newcommand{\ee}{\end{equation}}

\def\nn{\nonumber\\}
\def\calA{{\cal A}}
\def\calB{{\cal B}}

\begin{document}

\thispagestyle{empty}

\begin{titlepage}
\begin{flushright}
SU-ITP-2008-25, YITP-08-87\\
\end{flushright}


\vskip  2 cm

\vspace{24pt}

\begin{center}
{ \LARGE \textbf{ The footprint of $E_{7(7)}$ in amplitudes \\
of ${\cal{N}}=8$ supergravity  }}

\vspace*{0.6cm}

 {\Large
Renata Kallosh${}^{a,b}$   and Taichiro Kugo ${}^a$
} \\
 \vspace*{5mm}
 {\small
 ${}^a$ Yukawa Institute for Theoretical Physics, Kyoto 606-8502, Japan \\ \vskip2mm
 ${}^b$ Department of Physics, Stanford University, Stanford, CA 94305 \\
 }

\vspace{24pt}

\end{center}

\begin{abstract}

We study the low energy theorems associated with the non-linearly
realized continuous $E_{7(7)}(\mathbb{R})$ symmetry of the on-shell
${\cal{N}}=8$ supergravity.
For Nambu-Goldstone bosons we evaluate the one-soft-scalar-boson
emission amplitudes by computing the $E_{7(7)}$ current matrix element
on the one-particle external lines. We use the explicit form of the
conserved $E_{7(7)}$ Noether current and prove that all such matrix
elements vanish in the soft momentum limit,
assuming the $SU(8)$ symmetry of the S-matrix.
This implies that all tree amplitudes vanish in the one-soft-boson limit.
We also discuss the implications of unbroken $E_{7(7)}(\mathbb{R})$
symmetry for higher-order amplitudes.

\end{abstract}

\end{titlepage}
\newpage

\section{Introduction}

Classical ${\cal{N}}=8$ supergravity (SG) has, in addition to 8 local
supersymmetries, a local $SU(8)$ symmetry and a hidden global
$E_{7(7)}$ symmetry \cite{Cremmer:1978ds},\cite{deWit:1982ig} on shell,
when the exact non-linear equations of motion are satisfied. The
$E_{7(7)}$ symmetry is realized linearly and independently from the
local $SU(8)$ symmetry and it acts on 133 scalars present in the
classical action before gauge-fixing, as well as on the vectors of the
theory. The gauge-fixing can use the 63 local parameters of $SU(8)$ to
remove 63 non-physical scalars so that only 70 physical scalars are
left. This leads to a non-linear realization of the $E_{7(7)}$ on the
remaining 70 massless scalar fields. The $E_{7(7)}$ transformation has
to be performed simultaneously with the gauge preserving field dependent
$SU(8)$ transformation which was specified
 in \cite{Kallosh:2008ic}. In the light-cone gauge the first terms in
the coupling constant expansion of \E \, symmetry were recently
presented in \cite{Brink:2008qc}.

The purpose of this note is to study consequences of the
non-linearly realized $E_{7(7)}$ symmetry for the one-soft-scalar
emission. In our study of the low-energy theorems
\cite{Adler:1965ga,Weinberg:1996kr} in application to ${\cal{N}}=8$ SG
we will use the approach developed in \cite{Bando:1987br,Kugo:1981yf}
where the conserved current of the non-linearly realized ${G/H}$ symmetry plays
the major role. The corresponding Noether current was presented in
\cite{Kallosh:2008ic} following the procedure developed for the general
case of duality symmetries in \cite{Gaillard:1981rj}.

Recently Bianchi, Elvang and Freedman \cite{Bianchi:2008pu} were looking
for the footprint of $E_{7(7)}$ in tree diagrams of ${\cal{N}}=8$
SG.\footnote{The main results of the paper \cite{Bianchi:2008pu} is in a
construction of the generating functions for the \N=8 SG amplitudes and
their relation to \N=4 Yang Mills amplitudes.} The expectation was to
reveal the low energy theorems associated with the non-linear
realization of symmetries like in pion physics
\cite{Adler:1965ga,Weinberg:1996kr}. They have computed in
\cite{Bianchi:2008pu} the one-soft-boson limit of tree diagrams using
the Feynman rules and found that it always vanishes. Since there are
cubic interactions in the theory, the vanishing soft-boson limit of all
tree amplitudes was not an obvious feature of the theory but came out as
the result of careful computations. A different setting for the study of
the low-energy theorem was suggested in \cite{ArkaniHamed:2008gz} by
Arkani-Hamed, Cachazo and Kaplan. They used a specific supersymmetric
deformation of the \N=8 \, SG to complex momenta which provides a set of
recursion relations reducing all amplitudes to three point amplitudes.
This takes place due to a remarkable behavior of \N=8 \, SG at large
complex momenta. They studied  the 3-point amplitudes which
do not vanish on shell for the complex momenta.  They established that  in the
 one-soft boson limit these 3-point amplitudes vanish. This means, via  the recursion relations, that the one-soft boson limit for all on shell tree amplitudes vanishes.  They also studied a double soft limit of the amplitudes when two scalars are soft and found that it is related to an $SU(8)$ rotation of the amplitude without soft scalars.

In both of these references, \cite{Bianchi:2008pu} and
\cite{ArkaniHamed:2008gz}, the soft limit of the amplitudes are directly
computed and shown to vanish. Those amplitudes are related to the
low-energy theorem for an \E \, symmetry, but, by themselves, neither
prove nor disprove the \E \, symmetry.

In our approach here we will start with the \E \, symmetry and consider
the consequences of the Noether current conservation. The current $J_\mu$
consists of the linear part, proportional to the derivative of a scalar
$J_\mu^{\rm lin} $, and a non-linear part $J_\mu^{\rm nonlin} $, which starts as a
quadratic function of various fields. The total current conservation
relates the linear part to the non-linear part
\be
\partial^\mu J_\mu=\partial^\mu J_\mu^{\rm\, lin} + \partial^\mu J_\mu^{\rm\, nonlin}= 0 \ .
\label{cc}\ee
One can therefore derive the relation between the amplitudes
\be
  \langle\beta| \partial^\mu J_\mu^{\rm\, lin} | \alpha\rangle= -   \langle\beta| \partial^\mu J_\mu^{\rm\, nonlin}| \alpha\rangle \ .
\label{currentconserv} \ee
The left hand side of eq.  (\ref{currentconserv}) is related to the amplitude with the scalar emission ${\cal M} (\alpha\rightarrow\beta+\pi(k))$ since  $\partial^\mu J_\mu^{\rm\, lin}\sim\Box_x  \pi(x) $. When the scalar momentum $k$ is soft, the right hand side of the equation is proportional to amplitude without a scalar, where only the diagrams with singularities in the soft limit should be taken into account. The actual  computation in general \cite{Bando:1987br}  is reduced to the computation of the {\it divergence of the  non-linear part of the Noether current} between various one-particle states $ \langle i |$ and $| j \rangle$ divided by such singular propagator:
\be
g_A(0)_{ij} \equiv\lim_{k\rightarrow0} {\langle i (p)| \partial^\mu J_\mu^{\rm\, nonlin}(k)| j (p+k)\rangle\over p\cdot k} \ .
\label{KU}\ee
This expression was introduced in \cite{Kugo:1981yf} and
we will refer to it as `axial' charge, since it coincides with
the usual axial charge $g_A\,(\sim1.257)$ for the nucleon case in the pion
physics.
Clearly, there is one level of softness $k$ in the numerator due to the
factor $\partial$, but  there is a singularity from the propagator
in the diagram, which may cancel this $k$, and the soft limit of the
${\cal M} (\alpha\rightarrow\beta+\pi(k))$  may be non-trivial.
However, if $ \langle i | \partial^\mu J_\mu^{\rm\, nonlin}| j \rangle$ is
as soft as two powers in $k$,  the amplitude ${\cal M}
(\alpha\rightarrow\beta+\pi(k))$ will vanish in the soft limit.

 The discussion above is general and not restricted to any particular
level of perturbation theory. This is
{\em provided that} the linear part
$J_\mu^{\rm\, lin}$ is understood to be the part of the
current operator which gives the linear term not in the scalar field
but in the scalar {\em asymptotic}
field, that is, the whole part yielding the single massless scalar pole
terms (the category A diagram in
Fig.~1 below).
The full expression of the Noether current
is known \cite{Kallosh:2008ic}. Its matrix element between the
external one-particle states has to be computed in the soft-scalar limit to
establish the low-energy theorem. In this paper, however, we will limit
ourselves with the computation of the `axial' charge only at the tree
level.

First we will provide in Sec. 2 a calculational tool to derive the
low-energy theorem for theories with non-linear realization of a
symmetry with scalars in ${G}/{H}$ coset space. The low-energy theorem
in eq. (\ref{LET}) relates an arbitrary amplitude with an extra soft
scalar to the amplitude without such a scalar \cite{Bando:1987br}. The
relation between these amplitudes is defined by the `axial' charge
\cite{Kugo:1981yf}, which may or may not vanish, in general. In Sec. 3 we
explain the subtleties with the Noether current in \N=8 \, SG, which are
due to explicit appearance of the dual vector fields in the current.
These dual fields are not present in the Lagrangian. This would prevent us from
using the low-energy theorem in the form required for the analysis in
Sec. 2 based on \cite{Bando:1987br,Kugo:1981yf}. We show that if we
focus on a particular part of the
$E_{7(7)}$ current,
we may avoid this problem.
In Sec. 4 we actually compute various components of the `axial' charge
and show that they all vanish.
For this purpose we use only the quadratic in fields parts of the current.
The proof is generalized to whole $E_{7(7)}$ currents assuming
the $SU(8)$ symmetry of the S-matrix.
In Sec. 5 we discuss the steps towards the investigations of the
low-energy amplitudes in higher loop order.

\section{Low energy theorem for single pion emission processes}

Let us review the derivation of
the low energy theorem for single pion emission processes, following
Refs.~\cite{Weinberg:1996kr,Bando:1987br,Kugo:1981yf}.
We often call the Nambu-Goldstone (NG) bosons `pion' for short.
In our ${\cal N}=8$ SG context there are  {\bf 70} scalar particles.

The general setting is  that there is a symmetry group $G$ with a
continuous parameters $\epsilon_a$ where $a=1,\,\cdots,\,\dim\,G$.
In such a case there is a conserved Noether current
\be
\partial^\mu J_\mu^a=0.
\ee
Suppose that $G$ is spontaneously broken down to the unbroken subgroup $H$.
Then there are massless Nambu-Goldstone (NG) bosons $\phi^a$ in the
coset space ${G}/{H}$  whose number is equal to the
$\dim\, G- \dim \,H$,  i.e., $a=1,\,\cdots,\,\dim \,G- \dim \,H$.
The ``broken'' part of the  conserved Noether current has a linear term
as well as higher order terms nonlinear in fields
\be
J_\mu^a= -f^0_\pi\partial_\mu\pi^a + \cdots.
\label{eq:linearinJ}
\ee
One can sandwich the current between the vacuum and the one
NG boson state
\be
\langle0  | J_\mu^a(x) | \pi^b(k) \rangle= i k_\mu f_\pi\delta^{ab} e^{-ikx}
\label{J} \ee
where $f_\pi$ is the decay constant and is equal to $f^0_\pi$ in
(\ref{eq:linearinJ}) at tree level.
From the current conservation it follows that the NG boson is massless
\be
\langle0| \partial^\mu J_\mu^a(x) |\pi^b(k)\rangle=k^2 f_\pi\delta^{ab} e^{-ikx}
\qquad  \Rightarrow\qquad   k^2 =m_{\pi_a^2}=0
\ee
For the single soft pion processes we proceed as follows. Consider
the matrix element for emission of a single soft NG boson $\pi^a(k)$ in
an arbitrary multiparticle reaction $\alpha\, \rightarrow\, \beta$
\be
\langle\beta(P_f)\,{\out}  | J_\mu^a(x) | \alpha(P_i)\,{\iin} \rangle
\equiv M_\mu^a (k) _{\alpha\beta} e^{-ikx} \ , \quad P_i-P_f=k
\ee
As shown in Fig.~1, the diagrams contributing to this matrix element can be divided into
three categories according to the places where the current $J_\mu$ acts:
The first one (category A) includes those in which the current act at the
endpoint of the emitted pion line (which, therefore, come from
the field-linear term in $J_\mu$ at tree level.)
The second one (category B) includes those in which the current is attached
to an external line of the initial and final particles (which
come from the field-bilinear terms in $J_\mu$ at tree level.)
Finally, the third (category C) stand for the rest which develop no
one-particle pole singularity when $k^\mu\to 0$.
Following \cite{Bando:1987br} we represent these three contributions
as follows
\be
M_\mu^a (k) _{\alpha\beta}= {\cal P}_\mu^a (k) _{\alpha\beta}
+ {\cal Q}_\mu^a (k) _{\alpha\beta} + {\cal R}_\mu^a (k) _{\alpha\beta}.
\ee
\begin{figure}[hbt]
   \epsfxsize= 0.9\textwidth  
   \centerline{\epsfbox{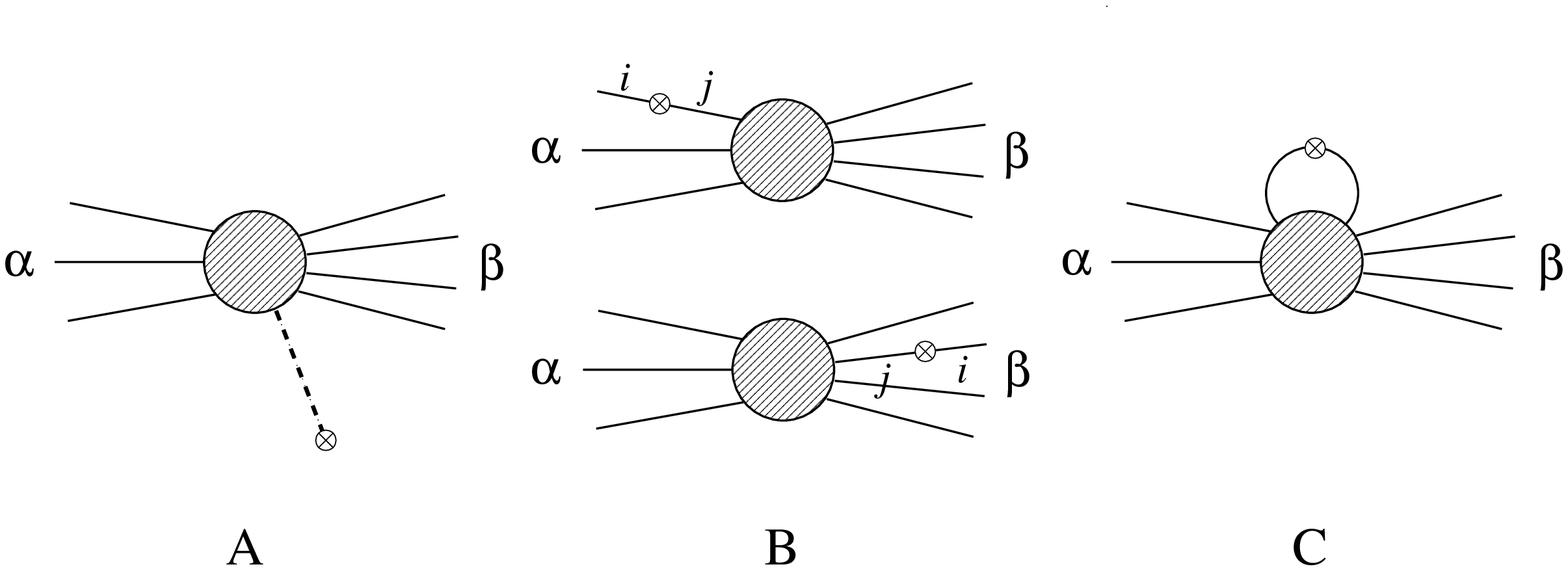}}
 \caption{Three categories of diagrams contributing to
$\langle\beta(P_f)\, {\out}  | J_\mu^a(x) | \alpha(P_i)\, {\iin} \rangle$.
The small circle with cross stands for the current operator. }
 \label{fig:1}
\end{figure}
The first term has a pion pole term of the form
\be
{\cal P}_\mu^a (k) _{\alpha\beta} e^{-ikx}= i f_{\pi} k_\mu e^{-ikx} {i\over k^2} G_{\alpha\beta}^\pi(k)
\ee
where $G_{\alpha\beta}^\pi(k)$ is the vertex function $\alpha\rightarrow\beta+\pi^a(k)$. For the
on-shell pion at $k^2=0$, $G_{\alpha\beta}^\pi(k)$ reduces to the physical
pion emission amplitude ${\cal M} (\alpha\rightarrow\beta+\pi^a(k))$ which we want to
compute:
\be
G_{\alpha\beta}^\pi(k)\Big|_{k^2=0}=i(2\pi)^4\delta^4(P_\beta+k-P_\alpha)
{\cal M} (\alpha\rightarrow\beta+\pi^a(k)).
\ee
Therefore, if we use the current conservation law $k^\mu M_\mu^a(k)_{\alpha\beta}=0$, we can evaluate the pion emission amplitude indirectly from the
other current matrix elements as
\begin{equation}
i(2\pi)^4\delta^4(P_\beta+k-P_\alpha)
f_\pi{\cal M} (\alpha\rightarrow\beta+\pi^a(k)) =
k^\mu\Bigl(
{\cal Q}_\mu^a (k) _{\alpha\beta} + {\cal R}_\mu^a (k) _{\alpha\beta}\Bigr).
\label{eq:qQ}
\end{equation}
If we are interested in the soft limit $k\ \rightarrow\ 0$ of
the amplitude, this implies that the only diagrams which have
singularities at $k=0$ can contribute to such soft pion amplitudes.

Such singularities can generally appear if the current operator
acts on the external one-particle lines as shown in the
diagrams of category B in Fig.~1, whose amplitude is denoted
by the second term $ {\cal Q}_\mu^a (k)_{\alpha\beta}e^{-ikx}$.
The rest diagrams in category C are regular at $k=0$
and cannot contribute. $ {\cal Q}_\mu^a (k)_{\alpha\beta}e^{-ikx}$
has two contributions, one when the current is attached to
the external line of the outgoing particle
\be
( {\cal Q}_\mu^a (k)_{\alpha\beta}e^{-ikx})_{\out}=
\sum_{i\in\beta_{\out}, m_j=m_i}
\langle i| J_\mu^a(x)|j\rangle^{\rm ext}\frac{i}{(p_i+k)^2-m_j^2}
\langle\beta-i+j|S|\alpha\rangle
\label{eq:Q1}
\ee
and the other, when the current is attached to the external line of the initial  particle
\be
({\cal Q}_\mu^a (k)_{\alpha\beta}e^{-ikx})_{\iin} =
  \sum_{i\in\alpha_{\iin}, m_j=m_i}  \langle\beta|S|\alpha-i+j\rangle\frac{i}{(p_i-k)^2-m_j^2}
\langle j| J_\mu^a(x)|i\rangle^{\rm ext}.
\label{eq:Q2}
\ee
Here $\langle i| J_\mu^a(x)|j\rangle^{\rm ext}$ denotes the external line
term ${\cal Q}_\mu^a (k)_{ij}e^{-ikx}$ for the single-particle case
$\alpha=i$ and $\beta=j$, and
$\alpha-i+j$ means that the particle $i$ in $\alpha$ is replaced by the
particle $j$ with the same momentum $p_i=p_j$ on the mass-shell.
It should be kept in mind that the `internal' states $|j\rangle$ and $\langle j|$ here
stand for slightly off-shell states before taking the soft limit $k\rightarrow0$, and
the expression
\be
\sum_j |j\rangle\frac{i}{(p_i\pm k)^2-m_j^2} \langle j|
\ee
should be understood to be the propagator of the particle $j$ (so that
the sum $\sum |j\rangle\langle j|$ over the polarization states gives the numerator of
the propagator like, e.g., $\slash{p}_j+m_j$ for the Dirac particle case.)

For the case where the particle $j$ is a massless gauge field, the
numerator $\sum |j\rangle\langle j|$ contain not only the physical transverse
states but also other unphysical polarization states. We will discuss
this point more concretely in Sect.4.

These external line terms (\ref{eq:Q1}) and (\ref{eq:Q2}) contain
 propagators which have singularities
$1/[(p_i\pm k)^2 - m_i^2] = 1/(\pm2p_i\cdot k+k^2)$ for on-shell momentum
$p_i^2=m_i^2$.

Now we can apply the current conservation $k^\mu M_\mu^a(k)=0$ and take
the limit $k\rightarrow0$. From Eqs.~(\ref{eq:qQ}) and (\ref{eq:Q1}),
(\ref{eq:Q2}), we find
\be
f_\pi{\cal M} (\alpha\rightarrow\beta+\pi^a(k)) \biggr|_{k\rightarrow0}=
i \left[ \sum_{i\in\beta,\ j} g_{A}^a(0)_{ji} {\cal M} (\alpha\rightarrow\beta-i+j)  -
 \sum_{i\in\alpha,\ j} g_{A}^a(0)_{ij} {\cal M} (\alpha-i +j \rightarrow\beta)  \right]
\label{LET}
\ee
where $g_A^a(0)_{ij}$ is an
`axial' charge of the external line
defined in \cite{Kugo:1981yf}
\be
g^a_{A} (0)_{ij} = \lim_{k\rightarrow0} \;
\frac{k^\mu{\cal Q}^a_\mu(k)_{ij}}{ 2p_i\cdot k \pm k^2}.
\label{eq:KUcharge}
\ee
This is the low-energy theorem for the single soft pion emission processes.

We  emphasize  here that we have only to evaluate the category B diagrams
in which the current is attached to the external lines thanks to the
current conservation. But we should note that this by no means implies
that
only the diagrams in which the pion couples to the external lines can
contribute to the soft pion emission amplitude. Indeed there are
generally diagrams in category A in which the pion
(denoted by dotted line) is attached
to the internal
lines/vertices but which give non-vanishing amplitude in the soft limit.

The diagrams in category B possess the one-particle singularity
so that the `charge' $g_A^a(0)_{ij}$ is generically non-vanishing.
However, in our ${\cal{N}}=8$ SG theory, no external lines
can give non-vanishing $g_A^a(0)_{ij}$ charge so that all the single
soft pion emission amplitudes vanish, as we will show below.

\section{$E_{7(7)}$ current of  ${\cal{N}}=8$ SG}

The classical non-gauge-fixed action of ${\cal{N}}=8$ SG has a
gauge $SU(8)$ symmetry with antihermitian and traceless local parameters
\be
\lambda_i{}^j (x)= -\lambda^j{}_i (x), \qquad \lambda^i{}_i (x) =0  , \qquad i=1,...,8.
\ee
and a global $E_{7(7)}$ symmetry with 133 parameters
\be
\epsilon_a = \{ \Lambda_I{}^J, \Sigma^{MNPQ}\} \qquad I, J, M, ...=1, ..., 8.
\ee
In the $E_{7(7)}$ symmetry we have generators of the $SU(8)$ maximal
subgroup of $E_{7(7)}$ with parameters $\Lambda_I{}^J$ and the orthogonal
ones, in ${E_{7(7)}}/{SU(8)}$ with parameters $\Sigma^{MNPQ}$.
There are 63 $\Lambda_I{}^J$  and they are antihermitian and traceless
\be
\Lambda_I{}^J= - \Lambda^J{}_I \qquad \Lambda^I{}_I=0\ .
\ee
They can be decomposed into 28 antisymmetric generators of the $SO(8)$
subgroup and 35 traceless symmetric generators orthogonal to $SO(8)$. If
we write $\Lambda_{I}{}^{J}$ as the sum of the real and imaginary parts $\Lambda
=\mbox{Re}\Lambda+i\, \mbox{Im}\Lambda$, then we have
\begin{eqnarray}\label{lambda12}
\mbox{Re}\Lambda^{\mathsf{T}}=-\mbox{Re}\Lambda\ ,\quad \mbox{Im}\Lambda^{\mathsf{T}}=\mbox{Im}\Lambda\ ,
\end{eqnarray}
where the real part is identified with the antisymmetric and the imaginary part with the symmetric part of $\Lambda$.
The off-diagonal part has to satisfy the self-duality constraint with the phase $\eta=\pm1$
\be
\Sigma_{IJKL}= {1\over24} \eta\, \epsilon_{IJKLMNPQ}  \Sigma^{MNPQ}\ .
\ee
We can also decompose $\Sigma$ into real and imaginary parts $\Sigma=\mbox{Re}\Sigma+i\, \mbox{Im}\Sigma$. However, in this case, both real and imaginary parts of $\Sigma$ have the same transposition properties
\begin{eqnarray}\label{sigma12}
\mbox{Re}\Sigma^{\mathsf{T}}=\mbox{Re}\Sigma\ ,\quad \mbox{Im}\Sigma^{\mathsf{T}}=\mbox{Im}\Sigma\
\end{eqnarray}
with $(\Sigma^{\mathsf{T}})_{IJKL}\equiv \Sigma_{KLIJ}$.
Then the self-duality constraint implies that the real part is $\eta
$-self-dual and imaginary part is $\eta$-anti-selfdual. The real and
imaginary parts of $\Sigma$ each consists of 35 real parameters. Thus we
present the 133 real parameters of $E_{7(7)}$ as $133=28+35+35+35$.

When the local $ SU(8)$ symmetry is fixed in the unitary gauge as described
in detail in \cite{Kallosh:2008ic}, there are only 70 scalars (out of
133) left in ${G}/{H}={E_{7(7)}}/{SU(8)}$. The Noether current
corresponding to $E_{7(7)}$ symmetry was presented in
\cite{Kallosh:2008ic} based on the general Gaillard-Zumino
procedure \cite{Gaillard:1981rj}.
\be
\partial^\mu J_\mu=0 \qquad  J_\mu\equiv
\sum_{a=1}^{133} J_\mu^a \epsilon_a
\ee
Here the 133 components of the current $J_\mu^a$ are contracted with the symmetry parameters $\epsilon_a$. 

The $E_{7(7)}$-current is special since it corresponds to the symmetry of
the equation of motion but {\em not} of the Lagrangian.
This peculiarity appears in the point that the current $J^\mu$
can be given only if we use the dual vector field $\calB_\mu$ which itself
does not appear in the lagrangian and is a complicated non-local
field if expressed in terms of the original fields, $\calA_\mu$ and others.

That is, the current is given in the form as given by
Gaillard and Zumino \cite{Gaillard:1981rj}:
\begin{eqnarray}
J^\mu&=& j^\mu_\varphi + j^\mu_{\rm GZ} \nn
j^\mu_\varphi&=&
\sum_{\varphi_i} {\partial{\cal L}\over\partial(\partial_\mu\varphi_i)} \delta^E\!\varphi_i \nn
j^\mu_{\rm GZ}&=&
{1\over4}(\tilde G^{\mu\nu}   A   {\cal A}_\nu
-\tilde F^{\mu\nu}   C   {\cal A}_\nu
+\tilde G^{\mu\nu}   B   {\cal B}_\nu
- \tilde F^{\mu\nu}   D   {\cal B}_\nu)\ ,
\label{eq:GZcurrent}
\end{eqnarray}
where $\varphi_i$ stand for all the fields other than the vector field
$\calA_\mu$, and $\delta^E\!\varphi_i$ for the $E_{7(7)}$ transformation of
$\varphi_i$. Here the $U(1)$ vector field strength $F$ and its dual $\tilde F$ are
$F= d {\cal A}$ and $\tilde F_{\mu\nu} \equiv{1\over2} \epsilon_{\mu\nu
\rho\sigma} F^{\rho\sigma}$, and $\tilde G_{\mu\nu}$ is defined to be
\begin{equation}
\tilde G^{\mu\nu} \equiv4{\partial{\cal L}\over\partial F_{\mu\nu}}\ ,
\label{eq:defGtilde}
\end{equation}
and the parameter matrices $A,\,\cdots,\,D$ are given by
\begin{equation}
A= -D^{\mathsf{T}} =\mbox{Re}\Lambda-\mbox{Re}\Sigma\qquad B=\mbox{Im}\Lambda+\mbox{Im}\Sigma
\qquad C=-\mbox{Im}\Lambda+\mbox{Im}\Sigma\,.
\label{eq:ABCD}
\end{equation}
If the equation of motion $\partial_\mu\tilde G^{\mu\nu}=0$ is used, its
dual $G_{\mu\nu}$ can be expressed as the field strength of the dual vector
field $G=d\calB$. Since $\tilde G^{\mu\nu}$ is
{\em defined} by
(\ref{eq:defGtilde}),  $G=d\calB$ is just an equation
of motion.

We have suppressed here the internal
indices on each vector field. In particular, in eq. (\ref{eq:GZcurrent})
the first two indices of $\Sigma^{ijkl}$ are contracted with the left vector
field strength ( $\tilde G$ or $\tilde F$ ) and the second two with the
corresponding two indices of the vector field ( ${\cal A}$ or ${\cal B}$ )
on the right.

The 4-divergence of Eq.~(\ref{eq:GZcurrent}) gives
\begin{eqnarray}
\partial j_\varphi &=& -\partial j_{\rm GZ} \nn
&=&-{1\over8}(\tilde G^{\mu\nu}   A   F_{\mu\nu} -\tilde F^{\mu\nu}  C   F_{\mu\nu}
+\tilde G^{\mu\nu}   B   G_{\mu\nu}
- \tilde F^{\mu\nu}   D   G_{\mu\nu}) \nn
&=&-{1\over8}( 2\tilde G^{\mu\nu}   A   F_{\mu\nu} -\tilde F^{\mu\nu}  C   F_{\mu\nu}
+\tilde G^{\mu\nu}   B   G_{\mu\nu} )
\ .
\end{eqnarray}
where we used $D^{\mathsf{T}}=-A$ in going to the last equality.

The problem here is the last term $\tilde G^{\mu\nu} B G_{\mu\nu}$, which
cannot be written in a 4-divergence form unless we introduce the
dual vector field $\calB_\mu$, even if we use the equation of motion.
Moreover, the Feynman rules in the presence of the dual
vector field $\calB_\mu$ are not available.
This
complicates the derivation of the low-energy theorem when this part of
the current is used.

Therefore, from here on, we restrict to the current (and soft scalars)
corresponding only to ${\rm Re}\Sigma$ of $E_{7(7)}/SU(8)$. We see from
Eqs.~(\ref{eq:ABCD}) that
only $A$ is non-vanishing and $B=C=0$ when $\Lambda={\rm Im}\Sigma=0$.
Thus we have
\begin{eqnarray}
&&J^\mu= j_\varphi^\mu- {1\over2}\tilde G^{\mu\nu}{\rm Re}\Sigma\,\calA_\nu
\end{eqnarray}
The explicit form for the $E_{7(7)}$ current $j^\mu_\varphi$ was given
in \cite{Kallosh:2008ic} and we will show the explicit forms for
the quadratic in fields part below where we will need them.

\section{Proof of the vanishing of the `axial' charge $g_A^a(0)$}

We now examine the scattering amplitudes for the single pion
emission processes (corresponding to ${\rm Re}\Sigma$).

Since the current $J^\mu$ is written solely in terms of the usual local
fields for the ${\rm Re}\Sigma$ cases (no dual vectors), it
is clear that only the singular diagrams for the current matrix element
$\langle\alpha|J^\mu(x)|\beta\rangle$ could contribute to the soft-limit
of the 4-divergence matrix element
\begin{equation}
\lim_{k\rightarrow0}\int d^4x \ e^{ikx}\,\langle\alpha|\partial_\mu J^\mu(x)|\beta\rangle.
\end{equation}
Those are the category A and B diagrams. As explained in Sect.2,
the pion emission amplitude given by the category A diagrams
can be evaluated by the category B
diagrams in which the current operator is inserted in the
external lines.

We thus have only to evaluate the category B diagrams.
For definiteness we consider the case where the soft pion is emitted with
momentum $k$ from the external lines appearing in the final states,
since the discussion for the initial state case is quite similar.
Then we want to evaluate the external line part as shown in Fig.~2.
\begin{figure}[htb]
   \epsfxsize= .5\textwidth  
   \centerline{\epsfbox{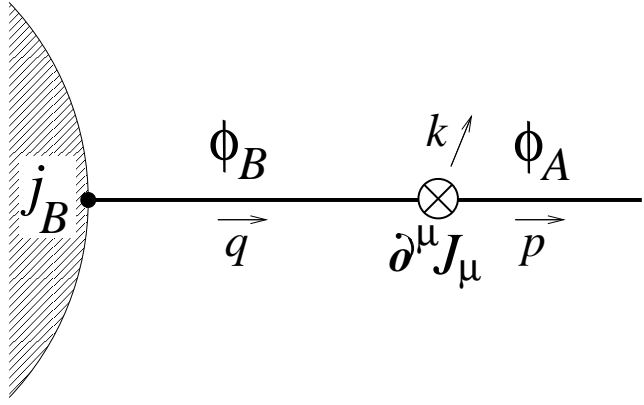}}
 \caption{The insertion of the non-linear part of the divergence of the Noether current into an external line of the on-shell amplitude.}
 \label{fig:1}
\end{figure}
We call the external on-shell state field $\phi_A(p)$,
and, looking back in time, it is converted by the current operator
into a slightly off-shell
particle $\phi_B(q)$ which propagates and connects to the source current
$j_B$ in the main body of
the diagram through the interaction Lagrangian $\phi_Bj_B$.
The external momentum $p$ and the pion momentum $k$ are put on the
mass-shell $p^2=k^2=0$ and we work in the frame in which
only $p^+$ and $k^-$ components are non-vanishing. Then the denominator
of the slightly off-shell propagator becomes
$q^2=(p+k)^2=2p\cdot k=2p^+k^-$.

Let us consider all case separately where those
$\phi_A(p)$ and $\phi_B(q)$ stand for various possibilities of fields.

\subsection{vector case; $\phi_A,\phi_B = \calA_\mu$}

The current of the vector field was given above and only
the quadratic part is relevant at the tree level:
\begin{eqnarray}
J^\mu_{\rm vec} &=& - {1\over2}\tilde G^{\mu\nu}{\rm Re}\Sigma\,\calA_\nu\nn
&\Rightarrow& - {1\over2}F^{\mu\nu}{\rm Re}\Sigma\,\calA_\nu.
\end{eqnarray}

Then, we separately evaluate the contributions from the two terms
$\partial_\mu\calA_\nu$ and $\partial_\nu\calA_\mu$ in $F_{\mu\nu}
=\partial_\mu\calA_\nu-\partial_\nu\calA_\mu$, since they are to be shown
vanishing separately.
The external vector line diagram with the current
$\partial_\mu\calA_\nu{\rm Re}\Sigma\,\calA^\nu$ inserted which is attached to the rest of the diagram through
the vertex $\calA_\rho j^\rho_V$ is evaluated as:
\begin{eqnarray}
&&-ik^\mu\left(\epsilon^*_\nu(p)\,ip_\mu{\rm Re}\Sigma\,{-i\delta^\nu_\rho\over q^2}
+\epsilon^{*\nu}(p){\rm Re}\Sigma\,(-iq_\mu){-i\eta_{\nu\rho}\over q^2} \right)
\bigl\langle j^\rho_V(q) \cdots\bigr\rangle\nn
&&\hspace{3em}{}=\epsilon^*_\nu(p)\,k\cdot(p-q)\,{\rm Re}\Sigma\,
{-i\delta^\nu_\rho\over q^2}\bigl\langle j^\rho_V(q) \cdots\bigr\rangle
= 0
\end{eqnarray}
since $q=p+k$ and $k^2=0$.

For the part $\partial_\nu\calA_\mu{\rm Re}\Sigma\,\calA^\nu$, we have
\begin{eqnarray}
&&-ik^\mu\left(\epsilon^*_\mu(p)\,ip_\nu\,{\rm Re}\Sigma\,{-i\delta^\nu_\rho\over q^2}
+\epsilon^{*\nu}(p)\,{\rm Re}\Sigma\,(-iq_\nu){-i\eta_{\mu\rho}\over q^2} \right)
\bigl\langle j^\rho_V(q) \cdots\bigr\rangle\nn
&&\hspace{3em}{}
=\left(k\cdot\epsilon^*(p)\,{\rm Re}\Sigma\,{-ip_\rho\over q^2}
-q\cdot\epsilon^*(p)\,{\rm Re}\Sigma\,{-ik_\rho\over q^2}\right)
\bigl\langle j^\rho_V(q) \cdots\bigr\rangle
= 0 \ ,
\end{eqnarray}
since the polarization vector is transverse so that
$k\cdot\epsilon^*(p)=k^-\epsilon^{*+}(p)=0$ and $p\cdot\epsilon^*(p)=0$
hence  $q\cdot\epsilon^{*}(p)=0$.

\subsection{fermion cases}

\def\slash#1{{\ooalign{\hfil/\hfil\crcr$#1$}}}

Next we consider the cases where $\phi_A(p)$ and $\phi_B(q)$ are the gravitinos
$\psi_{\mu i}$ and graviphotinos $\chi_{ijk}$, or vice versa.
The relevant current
operator at the tree level is the bilinear part:
\begin{equation}
J_{\rm ferm}^\mu\sim
\overline \chi_{ijk} \gamma^\nu\gamma^\mu\psi_{\nu l} \bar \Sigma^{ijkl} + h. c.
\label{FermionCurrent}
\end{equation}

\subsubsection{$\phi_A=\psi_{\mu i}$ and $\phi_B=\chi_{ijk}$}

The external gravitino should be on-shell
physical
so that the polarization vector-spinor $\psi_{(\pm)}^\mu(p)$
is of helicity $\pm3/2$:
\be \label{3/2state}
\psi_{(\pm)}^\mu(p) \equiv\epsilon_{(\pm)}^\mu(p)u_{(\pm)}(p)
\ee
where $\epsilon_{(\pm)}^\mu(p)$ is the transverse polarization vector
with helicity $\pm1$ and $u_{(\pm)}(p)$ is the Dirac spinor with helicity
$\pm1/2$. Note that this helicity $\pm3/2$ states satisfy the condition
$\gamma_\nu\psi_{(\pm)}^\nu(p)=0$, so that
$\bar \psi^\nu_{(\pm)}(p)\gamma^\mu\gamma_\nu= 2\bar \psi^\mu_{(\pm)}(p)$.
We use this relation in the fermion current (\ref{FermionCurrent})
and take the coordinate
system in which only $p^+$ and $k^-$ are
non-vanishing. Then, we find, for the external line part,
\be
-ik_\mu 2{\bar \psi^\mu_{(\pm)}(p)}\frac{i\slash{q}}{q^2}
=-ik^- 2{\bar \psi^+_{(\pm)}(p)}\frac{i\slash{q}}{2p^+k^-}
={\epsilon_{(\pm)}^{* +}(p)\,\bar u_{(\pm)}(p)}
\frac{\slash{q}}{p^+}
=0
\ee
since the transverse polarization vector $\epsilon^\mu_{(\pm)}(p)$
has vanishing $\mu=+$ components.

\subsubsection{$\phi_A=\chi_{ijk}$ and $\phi_B=\psi_{\mu i}$}

Next consider the case where gravitino is on
the slightly off-shell propagator side.
The gravitino propagator in the gauge with gauge-fixing term
\be
\frac{i}{2\alpha}(\bar\psi\cdot\gamma)\slash{\partial}(\gamma\cdot\psi)
\ee
is given by Das and Freedman\cite{Das:1976ct} in the form:
\begin{eqnarray}
\bigl\langle\psi_\nu(q)\psi_\rho(-q)\bigr\rangle
&=& i\frac{\displaystyle\bigl(\eta_{\nu\rho}+(2+\alpha){q_\nu q_\rho\over q^2}\bigr)
\slash{q}
+\frac12\gamma_\nu\slash{q}\gamma_\rho-(q_\nu\gamma_\rho+\gamma_\nu q_\rho)}{q^2} \nn
&=& i\frac{ -\displaystyle\frac12 \gamma_\rho\slash{q}\gamma_\nu+(2+\alpha){q_\nu q_\rho\over q^2}
\slash{q} } {q^2}
\label{prop2}
\end{eqnarray}
Writing the spinor state of graviphotino as
$\chi(p)$, we find
\be
\lim_{k\rightarrow0} -ik^\mu\
\overline \chi(p) \gamma^\nu\gamma_\mu
i\frac{ -\displaystyle\frac12 \gamma_\rho\slash{q}\gamma_\nu+(2+\alpha){q_\nu q_\rho\over q^2}\slash{q} } {q^2}
\bigl\langle j_\psi^\rho(q)\cdots\bigr\rangle
\ee
where $j_\psi^\rho$ is the source current of gravitino such
that $\bar\psi_\rho j^\rho_\psi$ appears
in the interaction part of the Lagrangian.
If we use the conservation law of the gravitino source current,
$q_\rho j_\psi^\rho(q)=0$, we immediately see that the double pole term vanishes.
This conservation law generally holds for the sum of a set of diagrams.
We can show that this double pole term actually vanishes graph by graph
as follows. Using the on-shell equation for the graviphotino $\overline
\chi(p)\slash{p}=0$, $\overline \chi(p)\slash{k}=\overline \chi
(p)\slash{q}$ and $k^2=0$, the double pole term is rewritten as
\be
k^\mu\
\overline \chi(p) \gamma^\nu\gamma_\mu
\frac{ q_\nu q_\rho\slash{q} } {(q^2)^2}
=
\overline \chi(p) \slash{q}\,\slash{k}
\frac{ q_\rho\slash{q} } {(q^2)^2}
=
\overline \chi(p) k^2
\frac{ q_\rho\slash{q} } {(q^2)^2}
= 0.
\ee
We can show that the rest part also vanishes as follows:
Taking the same coordinate system as above with only $p^+$ and
$k^-$ non-vanishing, and using the identity
$\gamma^\nu\slash{a}\slash{b}\slash{c}\gamma_\nu=
-2\slash{c}\slash{b}\slash{a}$,
\be
\lim_{k\rightarrow0} k^-\
\overline \chi(p) \gamma^\nu\gamma^+
\frac{ -\frac12 \gamma_\rho\slash{q}\gamma_\nu} {2p^+k^-}
=
\frac{-\frac12 \overline \chi(p) \gamma^\nu\gamma^+\gamma_\rho\slash{p}\gamma_\nu} {2p^+}
=
\overline \chi(p) \slash{p}\gamma_\rho\gamma^+
\frac{ 1} {2p^+} =0.
\ee

\subsection{scalar and graviton}

There are no scalar-scalar bilinear part in the current since the
scalar part of the current consists only of odd power terms in scalar
field (which is as usual in the non-linear Lagrangians for the
symmetric coset space $G/H$).
However there is a scalar-graviton bilinear term in the current:
\begin{eqnarray}
J^\mu_{\rm scalar-graviton} &=& \sqrt{-g}g^{\mu\nu}\,{\rm Re}\Sigma\,\partial_\nu y + h.c. \nn
&\Rightarrow& h^{\mu\nu}\,{\rm Re}\Sigma\,\partial_\nu y + h.c.
\end{eqnarray}
where we have defined our graviton $h^{\mu\nu}$ by
\begin{equation}
\sqrt{-g}g^{\mu\nu} = \eta^{\mu\nu} + \kappa h^{\mu\nu}
\end{equation}

\subsubsection{$\phi_A=h_{\mu\nu}$ and $\phi_B=y$}

For this case, the polarization tensor for the graviton external state is
given by the product of two polarization vectors for vector particle,
as $\epsilon^{\mu}(p)\epsilon^{\nu}(p)$, and the external
line part is evaluated as
\begin{equation}
-ik_\mu\, \epsilon^{*\mu}(p)\epsilon^{*\nu}(p)\,{\rm Re}\Sigma\,(-iq_\nu){1\over q^2}
=-k\cdot \epsilon^{*}(p)\ q\cdot \epsilon^{*}(p)\,{\rm Re}\Sigma\,{1\over q^2} .
\end{equation}
This vanishes since the polarization vector is transverse and
$p\cdot\epsilon^*(p)=0$, $k\cdot\epsilon^*(p)=0$ and
$q\cdot\epsilon^{*}(p)=0$ hold as for the above vector case.

\subsubsection{$\phi_A=y$ and $\phi_B=h_{\mu\nu}$}

In this case the external line diagram attached to the rest of the
diagram through the interaction term $h^{\rho\sigma}T_{\rho\sigma}$,
is given in the form
\begin{eqnarray}
-ik^\mu\cdot ip^\nu\cdot \,
i\, D_{\mu\nu, \rho\sigma}(q) \
\bigl\langle T^{\rho\sigma}(q) \cdots\bigr\rangle\ .
\label{grav}\end{eqnarray}
Here we use  the de Donder-Landau gauge for the graviton
$ \partial^\mu h_{\mu\nu}=0$ in which the
graviton propagator is given by\footnote{ See e. g.
\cite{Kallosh:1978wt} where the propagator for the graviton field
$\tilde h_{\mu\nu}$ ($ g_{\mu\nu} = \eta^{\mu\nu} + \kappa\tilde h_{\mu\nu}$) is given
in generic class of gauges. }
\begin{equation}
{\cal D}_{\mu\nu, \rho\sigma}(q) ={ \eta_{\mu\rho}\eta_{\nu\sigma}+ \eta_{\mu\sigma}\eta_{\nu\rho} \over q^2}-{     \eta_{\mu\rho} q_{\nu}q_{\sigma}+\eta_{\mu\sigma}  q_{\nu}q_{\rho}+
\eta_{\nu\rho} q_{\mu}q_{\sigma} +\eta_{\nu\sigma} q_{\mu}q_{\rho}\over q^4} +2 {q_{\mu} q_{\nu}q_{\rho}q_{\sigma}\over q^6}
\end{equation}
In eq. (\ref{grav}) the contribution from the second and third terms in the propagator immediately vanishes due to the explicit
factor of $q_\rho$ or $q_\sigma$ and because of the conservation of the energy-momentum tensor $q_\rho T^{\rho\sigma}(q)=q_\sigma T^{\rho\sigma}(q)=0$.
The first term vanishes in the soft pion limit
\begin{equation}
\lim_{k_\mu\rightarrow0}\,
{ 2k^-p_\mu\bigl\langle T^{+\mu}(q)\cdots\bigr\rangle\over2p^+k^-}=
{ p_\mu\bigl\langle T^{+\mu}(p)\cdots\bigr\rangle\over p^+} = 0
\end{equation}
because of the conservation of the energy-momentum tensor,
$p_\mu\langle T^{+\mu}(p)\cdots\rangle=0$.

We have thus completed the proof that the single soft pion emission
amplitudes vanish in ${\cal N}=8$ SG, at least for the soft
scalar particles ${\rm Re}\,y$ corresponding to the ${\rm Re}\Sigma$.
We cannot extend this proof directly to the scalars corresponding to
${\rm Im}\Sigma$ if we use the Feynman rules  from the action which has only one of the vector fields, not the dual one. This is because  the corresponding current cannot be
given without using the dual vector fields.
However, as Gaillard and Zumino argued, the Hamiltonian, and hence
$S$-matrix also, is invariant under $SU(8)$ transformation.
Since the scalar fields give an irreducible representation {\bf 70}
under $SU(8)$, we can conclude from $SU(8)$ symmetry of the $S$-matrix
that single soft pion emission amplitudes also vanish for the
${\rm Im}\,y$ scalar cases, once we prove that is the case for the
${\rm Re}\,y$ scalars. We should, however, keep in mind that the
$SU(8)$ symmetry is by no means trivial since it is not a manifest
symmetry in the Feynman graph computations but appears to be a symmetry of the on-shell amplitudes.

\section{$E_{7(7)}(\mathbb{R})$ symmetry in higher-loop orders?}

Before discussing the possibility of the higher-loop
$E_{7(7)}(\mathbb{R})$ symmetry with 133 generators $X$ and $T$, let us
remind that there are 70 generators of $ {E_{7(7)}/ SU(8)}$
symmetry, let us call them $X$, and there are 63 $T$-generators which
form the maximal $SU(8)$ subalgebra. The total algebra consists of $[T,
T]\sim T$ and $[X, T]\sim X$ and $[X, X]\sim T$.  At the tree level for
the amplitudes with any number of external states the following
information has been obtained at present. On one hand, the studies in
\cite{Bianchi:2008pu} and in \cite{ArkaniHamed:2008gz} were performed
directly on the amplitudes with an emission of a soft scalar and it has
been established that all such tree amplitudes vanish in the soft limit.
On the other hand, in this paper we have assumed that
$E_{7(7)}(\mathbb{R})$ is preserved and studied the consequences of such
assumption. One may argue that at the tree level the symmetry of the
on-shell action cannot be anomalous and therefore it is not even an
assumption that that $E_{7(7)}(\mathbb{R})$ is preserved in the form
\bea
&\partial^\mu J_\mu^{X} =(\partial^\mu J_\mu^{\rm lin})^{X} + (\partial^\mu J_\mu^{\rm nonlin})^{X}=0
 \eea
 for any  matrix elements between physical states at the tree level. We have computed the  amplitudes with an emission of a soft scalar associated with the term $(\partial^\mu J_\mu^{\rm lin})^{ X }$ indirectly by computing the matrix elements of the second term $(\partial^\mu J_\mu^{\rm nonlin})^{X}$. This second term could have provided us with the relation between the soft amplitude with a scalar and the amplitudes without a scalar as shown in eq. (\ref{LET}): the relation between these two is given by the `axial' charge  \cite{Bando:1987br,Kugo:1981yf}, which in our case is $g(0)_{ij}^{X}$. We have found that at the tree level in \N=8 \, SG all components of this charge are vanishing. Since at the tree level the conservation of the total Noether charge is taken for granted, $\partial^\mu J_\mu=0$, we have clearly an alternative derivation of the vanishing of soft amplitude with the emission of a boson. This follows from the $E_{7(7)}(\mathbb{R})$  symmetry.
 The subgroup $H=SU(8)$ of this symmetry just requires the current conservation and is not associated with any massless scalars
\be
\partial^\mu J_\mu^{T} = (\partial^\mu J_\mu^{\rm nonlin})^{T}=0\ .
\ee
At higher loop level we have to assume that the total Noether charge is
conserved, $\partial^\mu J_\mu$ both in the $SU(8)$ sector $T$ as well
as in the coset part of it, $X$. From such an assumption in the $X$ part
of the current we can only derive the low-energy theorem in the form of
eq. (\ref{LET}). By itself it does not require that the soft limit of
the amplitudes with a scalar should vanish: the symmetry only requires
that the soft limit is defined via eq. (\ref{LET}) to be related to the
amplitudes without a soft scalar times the `axial' charge.
One may entertain a scenario when at higher loops this charge is not
vanishing, and the soft limit of the amplitudes with a scalar is also
not vanishing. In such case   the  right hand side of equation (\ref{LET})
is equal to the left hand side, both non-vanishing,  and we may still have an unbroken $E_{7(7)}(\mathbb{R})$ symmetry.

However, if we look at the diagonal part of the $E_{7(7)}(\mathbb{R})$
algebra which is a $SU(8)$ subalgebra of it, the linear term is absent
since the scalars live in the coset space of $ {E_{7(7)}/SU(8)}$ and the
$SU(8)$ current has a usual structure of the type $\bar \psi\gamma_\mu
t^{IJ}\psi+...$ where the $ t^{IJ}$ matrices form the $SU(8)$ algebra.

The issue of the 1-loop anomalies is reduced
to the computation of the standard triangle anomaly diagrams. In \N=8 SG
this was done in \cite{Marcus:1985yy} where it was shown that $SU(8)$
anomalies cancel.
Anomalies for symmetries forming the algebra satisfy the  Wess-Zumino consistency condition.  Therefore
 the total $G=$\E \, may be anomaly-free and not only its maximum subalgebra.

What does this mean for the $E_{7(7)}(\mathbb{R})$ and the low-energy
theorems in higher order amplitudes? It is safe to expect that the
low-energy theorem (\ref{LET}), if confirmed, will prove that the coset part
of the symmetry, the $ {E_{7(7)}/SU(8)}$ part, is not anomalous. It is
also likely (but not necessary, from all we know) that it would mean
that the right hand side of eq. (\ref{LET}) vanishes by itself and the
left hand side by itself, i. e. the soft limit of the amplitude with a
soft scalar vanishes, as it takes place at the tree level.

In higher-loop level we have to find out if the low-energy theorem in
the form (\ref{LET}) is satisfied to preserve the $E_{7(7)}(\mathbb{R})$
symmetry. This requires both the knowledge of the one-soft scalar limit
amplitude as well as the computation of the `axial' charge at higher
loops. If the charge vanishes, as at the tree level and if the one-soft
scalar limit amplitude vanishes, the $E_{7(7)}(\mathbb{R})$ is
unbroken.

\section{Discussion}
 The second string revolution was, in particular, focusing on the U-duality of string theory, as explained in \cite{Hull:1994ys}, \cite{Witten:1995ex}.  It was noticed there that the $E_{7(7)}(\mathbb{R})$ symmetry of the classical ${\cal{N}}=8$ SG is broken down  by quantum effects to a discrete subgroup  $E_{7(7)}(\mathbb{Z})$ symmetry, which includes the T-duality group, $O(6,6, \mathbb{Z})$ and the S-duality group $SL(2, \mathbb{Z})$. It is a well known fact that the Noether theorem and the conserved Noether currents are associated  only with continuous symmetries and not with the discrete ones.

The relation between string theory and \N=8 SG in d=4 is not simple, moreover, it has been explained in \cite{Green:2007zzb} that the  perturbative  \N=8 SG in d=4   cannot be decoupled from the string theory. The reason for this is the existence in the string theory of additional massless and massive towers of states which are not present in  \N=8 SG in d=4. Therefore one should   study the  \N=8 SG  as a QFT, directly  in d=4. It has been even proposed that it may be the simplest possible QFT \cite{ArkaniHamed:2008gz}.

In this paper we studied the consequences of the classical continuous $E_{7(7)}(\mathbb{R})$ symmetry which leads to a conserved Noether current and explains why the one-soft-boson limit of all tree amplitudes  of ${\cal{N}}=8$ SG  vanishes.
Our method is complementary to the prior derivations of the low-energy
theorem in \N=8 SG in d=4. In \cite{Bianchi:2008pu}  it was found that all tree amplitudes
 vanish in the
one-soft-boson limit. This was established using
 the \N=8 SG Feynman rules and the hope was expressed that it might be
related to the $E_{7(7)}$ symmetry. We have now confirmed this and
clarified in the following sense: We have shown that the low-energy
theorem in \N=8 SG is a consequence of the continuous
$E_{7(7)}(\mathbb{R})$ symmetry, which remains unbroken as far as the
tree diagrams of \N=8 SG are concerned. The proof of the low-energy
theorem in \cite{ArkaniHamed:2008gz} supports the presence of the moduli
space in \N=8 \, SG. We derived the low-energy theorems associated
with the Nambu-Golsdtone bosons, coordinates of the ${G/H}=
{E_{7(7)}(\mathbb{R})/SU(8)}$ coset space and the corresponding
non-linearly realized symmetry.

Thus by now it is firmly established that
 the one-soft boson limit of all tree
amplitudes of \N=8 SG in d=4 vanishes.
 Moreover, it is shown here to be a direct consequence of the hidden $E_{7(7)}(\mathbb{R})$
symmetry of the perturbation theory.
 We use the generic form of the low-energy theorem for non-linearly
realized symmetries shown in eqs. (\ref{LET}), (\ref{eq:KUcharge}). From
this perspective the fact established in
\cite{Bianchi:2008pu,ArkaniHamed:2008gz} of the vanishing of the
one-boson-soft limit may not be sufficient to claim the symmetry, one
has to show in addition that the right hand side of the low energy
theorem in eq. (\ref{LET}) is also vanishing. In particular, the `axial'
charge, relating in general the soft limit to the amplitude without a
scalar may be vanishing. It is defined in eqs. (\ref{KU}),
(\ref{eq:KUcharge}). In the pion case this `axial' charge is
approximately $\sim 1.257$ for nucleon and therefore
the non-vanishing soft limit is
consistent with unbroken symmetry. We have computed the matrix elements
of the `axial' charge and have shown that it vanishes for the case of
$E_{7(7)}(\mathbb{R})$ symmetry in \N=8 SG.

The difference with \cite{Bianchi:2008pu,ArkaniHamed:2008gz} and the
value added by our work is the following. We have shown that generically
the symmetry requires that $ \langle\beta| \partial^\mu J_\mu^{\rm\,
lin} | \alpha\rangle= - \langle\beta| \partial^\mu J_\mu^{\rm\, nonlin}|
\alpha\rangle $ and we have specified both terms using the Noether
current. The detailed form of this equation is given in eqs.
(\ref{LET}), (\ref{eq:KUcharge}). The computation in
\cite{Bianchi:2008pu,ArkaniHamed:2008gz} of the one-soft-boson limit of
amplitudes with a scalar is the computation of the left hand side of
this equation. They found that $ \langle\beta| \partial^\mu J_\mu^{\rm\,
lin} | \alpha\rangle=0$. In our work we have established that the
right-hand side of this equation vanishes: we computed the matrix
element of the non-linear part of the Noether current and found that in
\N=8 SG $ \langle\beta| \partial^\mu J_\mu^{\rm\, nonlin}|
\alpha\rangle=0 $. Thus now we have a complete mechanism of the
manifestation of the $E_{7(7)}(\mathbb{R})$ symmetry at the tree level
in \N=8 supergravity: $ \langle\beta| \partial^\mu J_\mu^{\rm\, lin} |
\alpha\rangle$ was shown to vanish in
\cite{Bianchi:2008pu,ArkaniHamed:2008gz} and we have shown that $
\langle\beta| \partial^\mu J_\mu^{\rm\, nonlin}| \alpha\rangle $
vanishes. At the tree level one can assume that the total current is
conserved $\partial^\mu J_\mu=\partial^\mu J_\mu^{\rm\, lin} +
\partial^\mu J_\mu^{\rm\, nonlin}= 0$ and therefore one could have
computed either the first or the second term in the low energy theorem.
If one of them is vanishing, the other has to vanish. However, in case
when the symmetry will be studied beyond the tree level, the
conservation of the total current $\partial^\mu J_\mu=0$ should not be
taken for granted. The test of the $E_{7(7)}(\mathbb{R})$ symmetry
requires the knowledge of both terms in the current conservation
equation: $\partial^\mu J_\mu^{\rm\, lin} + \partial^\mu J_\mu^{\rm\,
nonlin}= 0$. It will not be sufficient to study only the soft limit of
the amplitudes, it will be necessary to compute the contribution from
the non-linear part of the Noether current as we have done it here via
the `axial' charge (\ref{KU}).

This brings us to the following question: Is the $E_{7(7)}(\mathbb{R})$
symmetry the property of tree diagrams only, or it will also take place
for higher order perturbation corrections? We presented an analysis of
this problem in Sec. 6. Finally, is $E_{7(7)}(\mathbb{R})$ symmetry
relevant to the issue of the conjectured all-loop finiteness of
${\cal{N}}=8$ SG \cite{Bern:2007hh},
\cite{Bern:2006kd}? This remains to be seen.

There is an argument  in favor of the absence of
anomalies of $E_{7(7)}(\mathbb{R})$ symmetry at the one-loop level. It
has been shown in \cite{Marcus:1985yy} that the chiral $SU(8)$ one-loop
triangle anomaly vanishes as a result of the cancelation of the
fermions and chiral vectors contribution. Since the $SU(8)$ has no
anomalies,  the Wess-Zumino consistency condition for anomalies
suggests that the total $G=$\E \, is not anomalous, at least at the
one-loop level.
 It would be very interesting to find out whether this
expectation is correct and study the status of possible \E \,
anomalies in higher loops.

\section*{Acknowledgments}

We are grateful to N. Arkani-Hamed, M. Bianchi, F. Cachazo, L. Dixon, T.
Eguchi, H. Elvang, S. Ferrara, D. Freedman, J.~Kaplan, M. Soroush, and
T. Rube for the most useful discussions of \N=8 supergravity. R. Kallosh
is grateful to the hospitality at Yukawa Institute of Theoretical
Physics where the main part of this work was performed. The work of RK
was supported by the NSF grant~ 0756174 and by YITP, Kyoto.
TK is
supported by the Grant-in-Aid for the Global COE Program "The Next
Generation of Physics, Spun from Universality and Emergence" from the
Ministry of Education, Culture, Sports, Science and Technology (MEXT) of
Japan. TK is also partially supported by a Grant-in-Aid for Scientific
Research (B) (No.\ 20340053) from the Japan Society for the Promotion of
Science.


\end{document}